\documentclass[aps,prl,reprint,superscriptaddress,nofootinbib]{revtex4-1}
\usepackage{graphicx}
\usepackage{amsmath}
\usepackage{amsfonts}
\usepackage{bm}
\usepackage{hyperref}
\usepackage[usenames,dvipsnames]{color}

\renewcommand\vec[1]{\mathbf{#1}}

\newcommand\abs[1]{|#1|}

\begin{document}
\frenchspacing

\title{Disconnecting structure and dynamics in glassy thin films}

\author{Daniel M. Sussman}\email{dmsussma@syr.edu}
\affiliation{Department of Physics, Syracuse University, Syracuse, New York 13244, USA}
\author{Samuel S. Schoenholz}
\affiliation{Google Brain, Google, Mountain View, California 94043, USA}
\author{Ekin D. Cubuk}
\affiliation{Department of Materials Science and Engineering, Stanford University, Stanford, California 94304, USA}
\author{Andrea J. Liu}
\affiliation{Department of Physics and Astronomy, University of Pennsylvania, Philadelphia, Pennsylvania 19104, USA}

\date{\today}

\begin{abstract}
Nanometrically thin glassy films depart strikingly from the behavior of their bulk counterparts. We investigate whether the dynamical differences between bulk and thin film glasses can be understood by differences in local microscopic structure. We employ machine-learning methods that have previously identified strong correlations between local structure and particle rearrangement dynamics in bulk systems. We show that these methods completely fail to detect key aspects of thin-film glassy dynamics. Furthermore, we show that no combination of local structural features drawn from a very general set of two- and multi-point functions is able to distinguish between particles at the center of film and those in intermediate layers where the dynamics are strongly perturbed.  
\end{abstract}

\maketitle


Confinement of glassy materials to nanometric length scales leads to striking changes to their microscopic dynamics and consequently to their material properties~\cite{Ediger2014}.  Direct observations in both experiments and simulations find exponentially more particle rearrangements near the free surface of a film~\cite{Fakhraai2008,Simmons2013,Scheidler2002, Hocky2014}.  A key question is whether enhanced dynamics near a free surface (or suppressed dynamics near a substrate) are connected with structural changes.  So far, all structural features studied decay too rapidly into the bulk~\cite{Ediger2014} to explain the altered dynamics. 

Until recently, the study of bulk glassy systems has also been plagued by the inability to connect dynamical changes with structural ones.  However, machine learning methods have proven remarkably successful in identifying a local structural quantity, termed ``softness" and denoted $S_i$ for particle $i$, that is strongly correlated with particle rearrangements~\cite{Cubuk2015,Schoenholz2016,SchoenholzArxiv}. Softness is over an order of magnitude more predictive of rearrangements than measures such as the local potential energy or coordination number~\cite{Cubuk2016}.  The average softness, $\langle S \rangle$, is directly predictive of the relaxation time of a bulk supercooled liquid~\cite{Schoenholz2016} or aging bulk glass~\cite{SchoenholzArxiv}, with higher values of $\langle S \rangle$ corresponding to shorter relaxation times at higher temperatures.  In bulk systems, it is therefore now clear that dynamical slowing down near the glass transition is intimately associated with structural changes.  Here we ask whether the enhancement of dynamics near the surface of free glassy films can similarly be understood. 

The answer is no.  Not only does softness fail to predict the enhanced dynamics near the surface of free glassy films, we find that for a very general set of quantities that characterize the local structural environment surrounding a particle, there is no combination of these quantities that can distinguish between parts of the film with very different dynamics. The enhanced dynamics near a free glassy surface therefore appear to be fundamentally different from the enhanced dynamics that result from heating bulk glassy systems.  Although we cannot rule out the possibility that structural quantities that we have not considered might tell a different story, our results suggest that near glassy free surfaces, relaxation is dominated by mechanisms that are independent of local structure.


Our model systems are composed of short-chain polymers, specifically Kremer-Grest chains each with $N=20$ beads of diameter $\sigma$. The non-bonded interaction is an attractive, truncated Lennard-Jones (LJ) potential and the intrachain bonds are stiff harmonic springs.  This model has a bulk glass transition temperature of  $T_g=0.44\pm 0.01$. We prepared both freestanding films and films on frozen amorphous substrates with thickness of order $30\sigma$, according to the protocols in Refs.~\cite{Shavit2014,Sussman2016}.  Here, $\hat{z}$ is normal to the film surface. The systems were simulated using the HOOMD-blue package \cite{hoomd-blue,hoomd-blue0} in the NVT ensemble with a time step of $\delta t=0.001\tau$, where $\tau$ is the LJ unit of time.  All configurations were prepared at an initial reduced temperature of $T=0.8$ and cooled at a constant rate of $\Gamma = 1\times 10^{-3}$ to the desired temperature; the aging time $t_{ag}$ refers to simulation time subsequent to the end of this cooling procedure. Additional details can be found in the Supplemental Material~\cite{SupMat}.

A connection between a particle's local structural environment and its propensity to rearrange has been established in bulk glasses in the last several years~\cite{Berthier2007,Reichman2008,Royall2008,Manning2011,Schoenholz2014}. We follow the machine learning work of Ref.~\cite{Schoenholz2016} and study $S$.  We begin by recording particle trajectories within $5\sigma$ of the center of mass of an aged thin film at $T=0.425$. We use a ``hop'' indicator function, $p_h(i;t)$, to identify persistent particle motions and rearrangements \cite{Candelier2009}. To define $p_h(i;t)$ for a particle $i$ at time $t$ we first specify two time intervals $A=\left[ t-5\tau,t\right]$ and $B=\left[t, t+5\tau\right]$; the hop indicator function can then be expressed as
\begin{equation}
p_{h}(i;t)=\sqrt{  \langle \left( \mathbf{r}_i - \langle \mathbf{r}_i\rangle_B\right)^2\rangle_A \langle \left( \mathbf{r}_i - \langle \mathbf{r}_i\rangle_A\right)^2\rangle_B  },
\end{equation}
where $\langle\rangle_A$ and $\langle \rangle_B$ denote averages over $A$ and $B$ intervals. From the $p_h$ trajectories of the central particles, we identify a ``training set" of $2000$ particles that are about to rearrange, such that $p_h > p_c = 0.2$ in the subsequent frame, and 2000 particles that have not rearranged for a long time, where $p_h < p_{c,l} = 0.007$ for many hundred $\tau$. 
We note, however, that the definition of $p_h$ could obscure the contribution of smoother particle motions to structural relaxation. We characterize the local environment of each particle in the combined set using $M$ ``structure functions,'' $G_\alpha$. In Ref. \cite{Schoenholz2016} we showed that we could limit our structure functions to local, coarse-grained versions of the pair correlation function:
\begin{equation}
G_r(i;r,\sigma) = \frac{1}{\sqrt{2\pi}} \sum_{j\in \mathcal{X}} e^{-(R_{ij}-r)^2)/2\sigma^2},
\end{equation}
where $R_{ij}$ is the distance between particles $i$ and $j$, $\mathcal{X}$ is the set of particles near particle $i$ with $R_{ij}<2.5\sigma$, and $r \in \left[0.9, 1,1.1,\ldots,2.5\right]$. 

We embed the particles in our training set in the space of structure functions, $\mathbf{G}_i$, and find the hyperplane in $\mathbb{R}^M$ that best separates the two classes. The softness is the signed distance to the plane: $S_i = \mathbf{w}\cdot\mathbf{G}_i + b$, where $\mathbf{w}$ is the hyperplane normal and $b$ is the bias. One notable difference in protocol compared to Ref.~\cite{Schoenholz2016} is that here we use snapshots of thermalized configurations of the particles, rather than quenching to their inherent structures (thus avoiding spurious rearrangements due to thermal contraction from the thermalized to the quenched state of the film). Nevertheless, we find that our predictive accuracy remains high, with approximately $84\%$ of rearranging particles identified as soft (i.e., with $S_i > 0$), in comparison to the $88 \%$ accuracy found using radial structure functions on the inherent structures of bulk systems~\cite{Schoenholz2016,Cubuk2016}. We have verified that our qualitative conclusions are insensitive to choices made, such as using additional structure functions (bond-angle dependent functions~\cite{Cubuk2015} or spherical harmonics coarse-grained in the radial direction), or training on bulk particles at the same temperature and density as the thin films.


\begin{figure}[htbp]
\centerline{\includegraphics[width=0.45\textwidth]{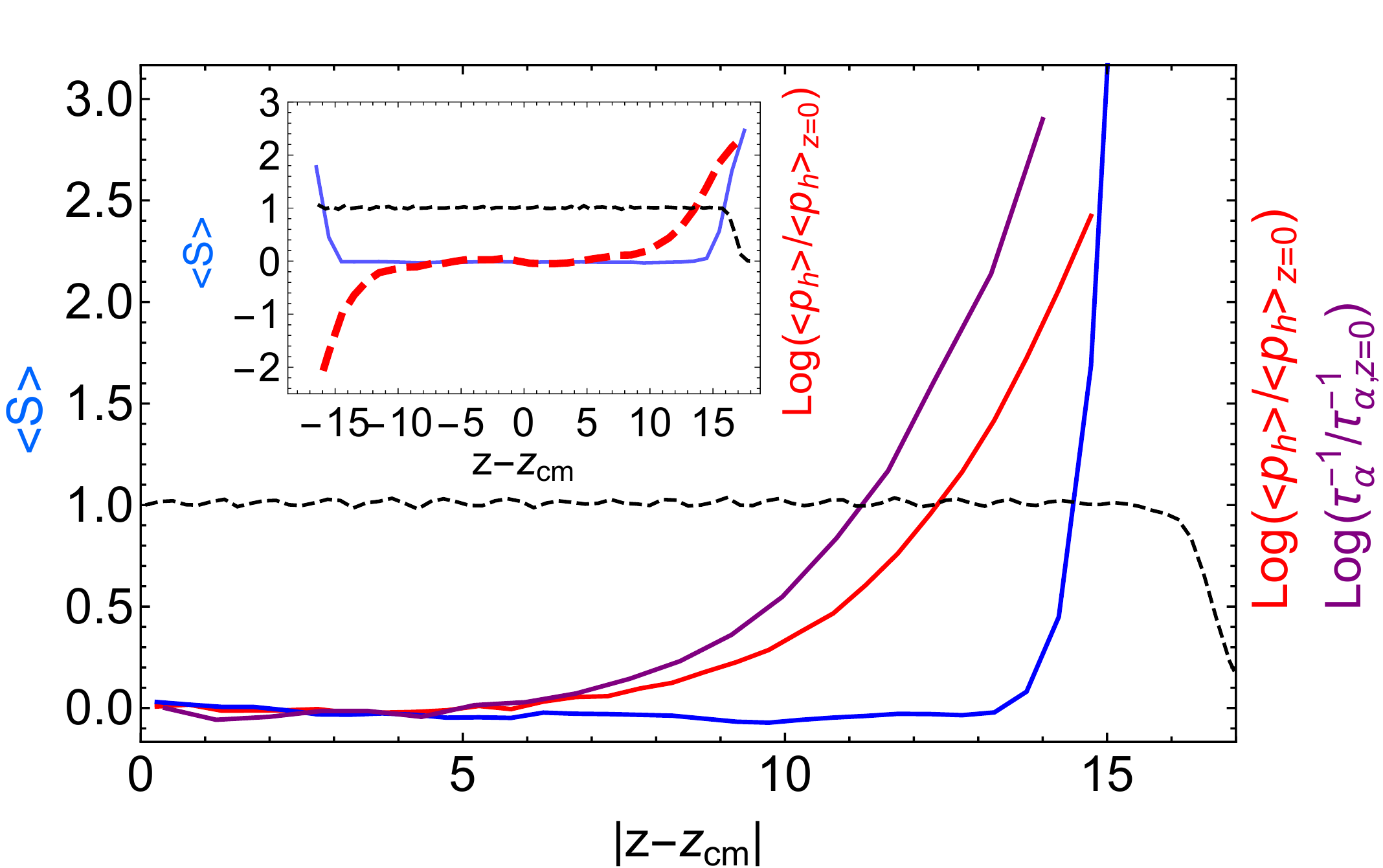}}
\caption{\label{fig:freefilmt} 
Spatial profile of mean softness (blue, lowest curve), log of normalized $\langle p_h \rangle $ (red, intermediate curve), log of normalized inverse relaxation time (purple upper curve), and normalized density profile (black, dashed) for free standing thin film at $T=0.425$ and $t_{ag} \approx 5 \times 10^5$ ($\tau_{\alpha,z=0}\approx 2.2\times10^4$). Inset. Spatial profile of mean softness (blue, solid curve), log of normalized $\langle p_h \rangle$ (red, dashed curve), and normalized density profile (black, dashed curve) for a supported film at $T=0.45$ and $t_{ag} \approx 6 \times 10^5$. ($\langle p_h\rangle_{z=0}\approx 0.27$)}
\end{figure}

Using these methods we compare the dynamical properties of thin films with spatially-resolved values of particle softness. Figure \ref{fig:freefilmt} shows the layer-resolved average value of the softness field, and the logarithms of the mean value of the $p_h$, and the $\alpha$ relaxation time for each layer (as estimated from the decay of the overlap function \cite{Keys2007}). To put the dynamical properties on a comparable scale, we normalize them by their values at the center of the film. Strikingly, the mean value of softness is flat over nearly the entire film.  In particular, we call attention to the regime between $\abs{z-z_{\textrm cm}}=7$ and  $\abs{z-z_{\textrm cm}}=13$, where the mean softness is flat even while the dynamics are speeding up by an order of magnitude. The value of softness only changes for particles close enough to the edge of the film for $G_r$ to detect the density fluctuations at the interface.  At a distance of within $2.5\sigma$ of the surface, the values of the structure functions are affected by the fact that particles are missing beyond the edge of the film; at this point $S$ must lose its predictive value. To provide a sense of scale, we note that relaxation times in Kob-Andersen (KA) LJ mixtures in both supercooled and aging bulk systems are well-described by
$\tau_\alpha (T) \sim \exp \left[ \left(\alpha_1 \langle S \rangle -\alpha_2 \right) \left(T_0^{-1}-T^{-1} \right)   \right],$
where $\alpha_1$ and $\alpha_2$ are temperature-independent parameters and $T_0$ is the dynamical onset temperature  \cite{SchoenholzArxiv}. Thus, in a bulk system the differences in $\tau_\alpha$ or  $\langle p_h \rangle$ as large as those observed between, e.g., the center of the film and $|z-z_{cm}|=10$ would cause changes in the mean softness much larger than the numerical fluctuations observed in this data.

The inset of Fig. \ref{fig:freefilmt} reports results for a film at $T=0.45$ that is supported by an amorphous substrate. Here the dynamics are exponentially suppressed near the solid interface and enhanced at the free surface.  Note that if softness were predictive of the dynamics, we would expect the average softness to decrease near the solid substrate, and to increase near the free surface. In contrast to this expectation, the value of $\langle S\rangle$ shows no indication of a difference in the local structural environment of the particles; again it remains flat except within $2.5 \sigma$ of the surfaces. To treat both sides symmetrically, we do not include substrate particles in calculating the structure functions; we thus see an upturn in softness associated with the loss of particle density on both sides.  Note that amorphous walls have been used to detect point-to-set-like length scales ~\cite{Kob2012,Hocky2014}, but these structural length scales are short enough ($\sim \sigma/2$ in Kob-Andersen mixtures) to decay inside the $2.5\sigma$ range of the structure functions, and are thus buried by the effects of density loss for particles near the substrate.

\begin{figure}[htbp]
\centerline{\includegraphics[width=0.45\textwidth]{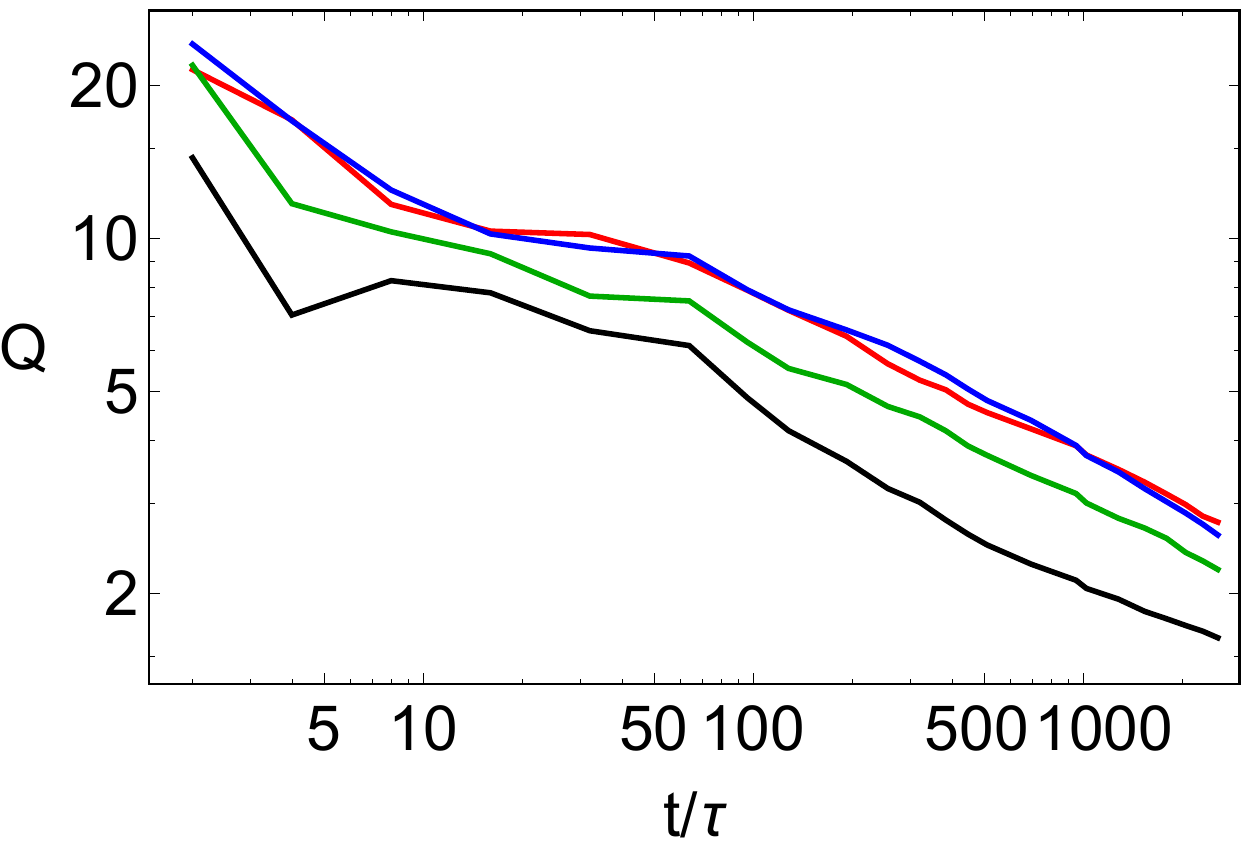}}
\caption{\label{fig:qplot} 
$Q(t)$ for freestanding thin film at $T=0.425$ and $t_{ag} \approx 2 \times 10^5$. From top to bottom, the curves are averages over particles with $|z-z_{cm}| < 4\sigma$, $4\sigma < |z-z_{cm}| < 8\sigma$ , $8\sigma < |z-z_{cm}| < 10\sigma$ , $10\sigma < |z-z_{cm}| < 12\sigma$.}
\end{figure}

We find that all of our supercooled and aging film systems, both freestanding and supported, display this insensitivity of the structural features to film position, and by extension to the changes in the dynamics. Does softness, then, maintain any predictive power in the intermediate layers of the film? We quantitatively answer this by using the $Q$-function proposed in Ref. \cite{Cubuk2016} for an arbitrary structural quantity, $X$:
\begin{equation}
Q_X(t) = \frac{P_R\left(X_i > \mu_X + \sigma_X \right)}{P_R\left(X_i < \mu_X - \sigma_X \right)},
\end{equation}
i.e., the ratio of the probability of rearrangement for particles with measure $X$ a standard deviation above and below the mean, as a function of time since $X$ was measured. For context, we report $Q(2\tau)$ for inherent structures of bulk KA mixtures at $T=0.47$ and density $\rho=1.2$ using different features $X$ to try to predict rearrangements: $Q_z=6.2$ where $z$ is the coordination number, $Q_E=5.0$ where $E$ is the local energy, and $Q_S=165$ where $S$ is softness \cite{Cubuk2016}. Figure \ref{fig:qplot} reports $Q(t)$ for softness evaluated on several different layers in a thin film. We note that $Q$ decays in a nearly-scale free way over more than three decades of time, a range that covers roughly an order of magnitude less than the relaxation time at the center of the film. Even in the intermediate layers of the film, $S$ retains a surprisingly high predictive power well beyond the $2\tau$-time scale rearrangements that the softness field was trained on. However, as one moves away from the center of the film $Q$ shifts downwards, indicating that the overall ability of softness to predict rearrangements decreases.  

Taken together, the results above imply that the strong connection between structure and dynamics in bulk systems breaks down in the presence of an interface. We therefore abandon our efforts to correlate structure with local dynamics. Instead, we ask the simpler question of whether the local structural environment of a particle depends \emph{in any way} on the particle's position in the film.  To answer this, we directly perform supervised learning on the position of a particle using SVM with linear and RBF kernels as well as feedforward neural networks with up to six fully connected layers (see Supplemental Material for details \cite{SupMat}), focusing on the freestanding films described above ($T=0.425$ and $t_{ag}\approx 5\times 10^5 \tau$), and take $z_{cm}=0$. Instead of training on particles that are rearranging or not, we construct a training set consisting of ``central'' particles (with $|z| < 1.25$) and ``edge'' particles (with $11.0 < |z|  <12.25$, i.e. close to the edge but still more than $2.5\sigma$ from the density fluctuations near the film boundary).   The relaxation time in the center layer is $\tau_{\alpha}\approx 2.2\times10^4$ while that in the edge layer is $\tau_{\alpha,z=0}\approx 6.5\times10^3$. Can we distinguish between these two sets of particles using local structure alone?

In order to capture many different aspects of the local structural environment, we expand our set of structure functions beyond the purely radial ones used above. We include isotropic spherical-harmonic bond orientational order functions, $G_{Q_l}$ \cite{Steinhardt1983,Schoenholz2016}. We also employ a collection of spherical harmonics, $G_{Y_{lm}}$, using $z$ as the axis of symmetry. This is potentially quite important, as the structure functions should reflect the symmetry of the underlying system itself. We compute the $G_{Q_l}$ and $G_{Y_{lm}}$ for even $l$ (up to $l_{max}=12$) in shells of thickness $\sigma/2$; explicit definitions are given in the Supplemental Material \cite{SupMat}. Training on 2500 particles from each class, we find cross-validation accuracies of $50\%$; that is, we are completely unable to distinguish particles in different layers of the film using this very general set of radial and angular structure functions. 

One possible source of the altered dynamics near an interface lies in changes not to the mobility of particles but to their 
mobility-mobility correlations. As a first effort to detect these subtler effects, we introduce another new class of structure functions. Since in the bulk softness itself is a good predictor of mobility, we first compute the softness of every particle as described above. We then compute the local softness densities at different distances from each particle:
\begin{equation}
G_S(i;r,\sigma) = \frac{1}{\sqrt{2\pi}} \sum_{j\in \mathcal{X}} S_j e^{-(R_{ij}-r)^2)/2\sigma^2}.
\end{equation}
This captures radial features of the softness field itself. This slightly increases our predictive accuracy, but nevertheless fails to distinguish positions within the film. While structure functions of this sort are thus not useful in understanding thin film dynamics, we speculate that adding these functions to the usual set in the bulk might allow generalized versions of softness to predict the intermediate time and length scale phenomena associated with dynamical heterogeneities.

\begin{figure}[htbp]
\centerline{\includegraphics[width=0.4\textwidth]{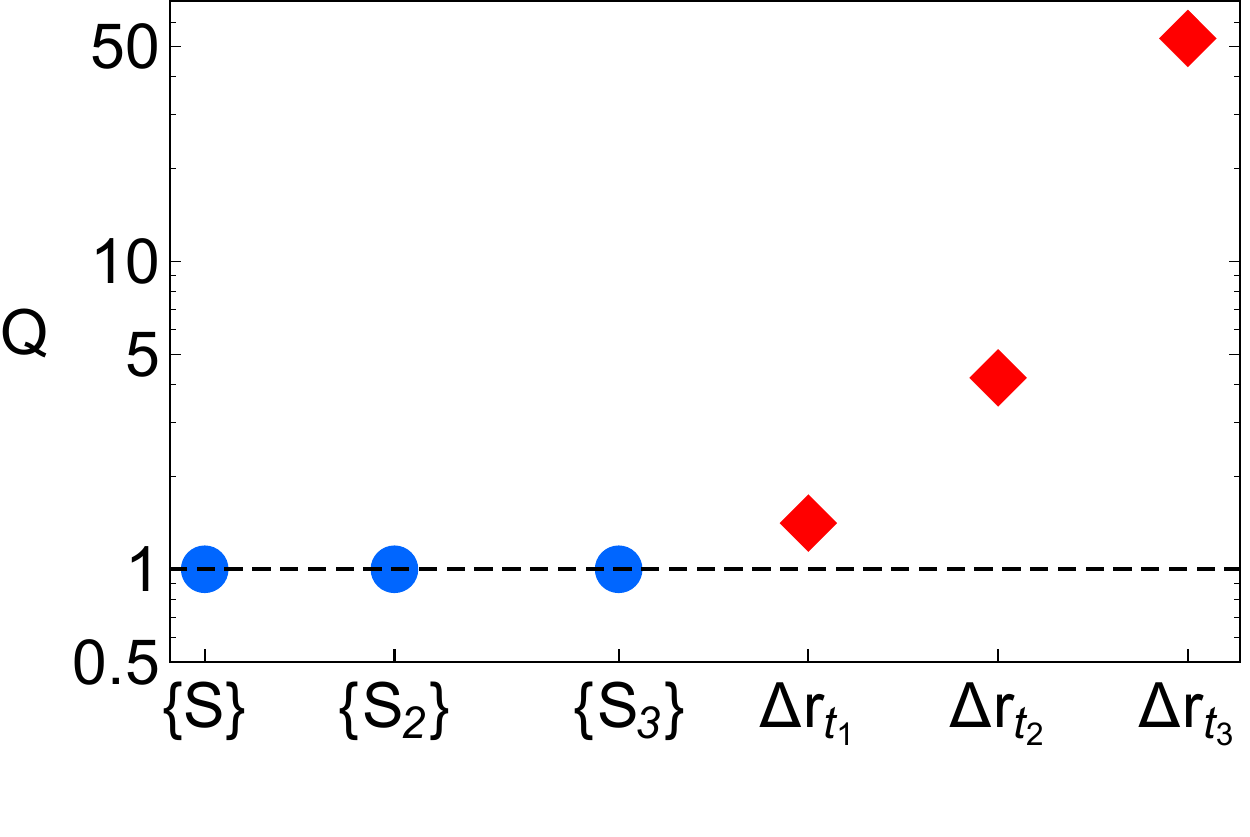}}
\caption{\label{fig:scoreplot} 
$Q$ function scores obtained trying to distinguish 2500 particles near the center versus the edge of the thin film, as a function of the set of structure functions used. The three circles correspond to using only local structural information: $\{S\}$ are the standard softness, $\{S_2\}=\{S,G_{Q_l},G_{Y_{lm}}\}$, $\{S_3\}=\{S_2,G_S\}$.The three diamonds correspond to adding dynamical information (logarithmically spaced data on the magnitude of particle displacement, $\Delta r_i$, out to a maximum time of $t_i = 2\tau, 200\tau, 2000\tau$, as described in the Supplemental Material~\cite{SupMat}.}
\end{figure}

In contrast, we stress that \emph{dynamical} information can predict position within the film, as we would expect. Adding structure functions that correspond simply to the average magnitude of particle displacement at logarithmically sampled times less than $t$, allows increasingly accurate predictions of particle position as the maximum time interval grows. (The prediction accuracy should level off once the time interval $t_i$ is comparable to the relaxation time in the edge layer \cite{SupMat}.) The results for the predictive power of these structural and dynamical features, again in terms of their respective $Q$ functions, is summarized in Fig. \ref{fig:scoreplot}.


In conclusion, we have applied machine learning methods to show that there is no detectable correlation between structure and enhanced or suppressed dynamics near the surfaces of thin glassy films. We showed that softness -- which has been highly successful in capturing many important features of glassy dynamics in bulk systems~\cite{Cubuk2015,Schoenholz2016,SchoenholzArxiv,Cubuk2016} -- utterly fails to predict changes in the dynamics as a function of position within the film. We further showed that a supervised learning algorithm fails to find any difference in the local structure at the center of the film compared to the edge.  This is true even after radically enlarging the space of local structure quantities we consider to very general classes of two- and multi-point quantities. Although we cannot consider \emph{all} possible structural quantities, our results strongly suggest that the local structure is the same everywhere except very close to the edge of the film.  The enhanced/suppressed dynamics at the free/supported surfaces of thin glassy films therefore appear to be fundamentally different from those that dominate the relaxation of bulk glassy systems.

The behavior of thin glassy films has often been interpreted in terms of a two-population model, in which there is a glassy, immobile layer near the center of the film and a liquid-like mobile layer near the free surfaces \cite{Ediger2014,Tsui2010,Meyers1992,Chai2014}. This interpretation is supported by observations of probe molecules embedded in films \cite{Ediger2011,Fakhraai2014}.  Our results show that the two populations are indistinguishable from a structural point of view.

Instead, we suggest that the two populations are distinguished by different mechanisms of relaxation.  In the immobile layer near the center of the film, rearrangements preferentially involve particles with certain local structural environments (high softness).  The enhancement of relaxation near the free surface of a glassy surface (or the suppression in the case of a supported surface) is completely unpredicted by softness and therefore relies on a different mechanism. Existing theoretical models of the behavior of thin glassy films start from a diverse set of assumptions, considering configurational entropy, facilitated dynamics, and the effect of changing elastic moduli on the barriers to local rearrangements~\cite{Wolynes2008,Starr2014,Harrowell2010,Chandler2010,Milner2013,Mirigian2015}. Our results strongly favor theoretical descriptions that do not rely on structural differences to propagate the effect of the interface into the film.

In bulk systems the probability of a particle rearranging at a given value of softness is well-described by \begin{equation}\label{eq:probability_rearrange}
P_R(S) = \exp\left(\Sigma_0 - e_0/T\right)\exp\left[ S \left(-\Sigma_1 + e_1/T\right)\right],
\end{equation}
where the constants $\Sigma_0,\ \Sigma_1, \ e_0,\ e_1$ are independent of $T$ (with non-Arrhenius behavior arising from the dependence of the $S$-distribution on temperature) \cite{Schoenholz2016}. Thus, $P_R(S) = P_I(T) P_D(S,T)$ where $P_I(T)$ and $P_D(S,T)$ are structure-independent and -dependent contributions to $P_R(S)$, respectively. This form suggests that rearrangements in glassy liquids occur when two uncorrelated processes (one that depends on structure and one that does not) coincide. (If these processes could \emph{independently} cause rearrangements then we would expect $P_R(S) = P_I(T) + P_D(S,T) - P_I(T) P_D(S,T)$.) In the film, interfaces appear to primarily affect particle rearrangements through $P_I$ \cite{SupMat}, although more detailed studies of enhanced particle mobility near surfaces are required to disentangle the two processes and to determine whether the mechanism underlying $P_I$ in bulk systems is the same as that near free surfaces. 

\begin{acknowledgments}
We thank Robert Riggleman and Amit Shavit for helpful discussions and for sharing data on their thin film configurations. This work was supported by the Advanced Materials Fellowship of the American Philosophical Society (DMS), the UPenn MRSEC DMR-1120901 (D.M.S. and A.J.L.), the U.S. Department of Energy, Office of Basic Energy Sciences, Division of Materials Sciences and Engineering under Award DE-FG02-05ER46199 (S.S.S. and A.J.L.), and the Simons Foundation (327939 to A. J. L.). The Tesla K40 used for this research was donated by the NVIDIA Corporation.
\end{acknowledgments}

\bibliography{surface_softness_bib}

\begin{thebibliography}{36}
\bibitem [{Sup()}]{SupMat}%
  \BibitemOpen
  \href@noop {} {\bibinfo  {journal} {Suplemental Material can be found at
  [link to be inserted by publisher]}\ }\BibitemShut {NoStop}%
\end{thebibliography}

\newpage

\appendix
\section{Supplemental Material}

\subsection{Simulation details}
We perform molecular dynamics simulations of a commonly used coarse-grained model polymer \cite{Grest1990} prepared in thin film geometries. Our systems have $N_{total}=80000$ total particles of diameter $\sigma$, composed of monodisperse chains of length $N=20$. The non-bonded interactions between particles $i$ and $j$ are specified by 
\begin{equation}
V_{ij}^{nb}=4\epsilon \left( \left( \frac{\sigma}{r_{ij}} \right)^{12} - \left( \frac{\sigma}{r_{ij}} \right)^6 \right) -4\epsilon \left( \left( \frac{\sigma}{r_{c}} \right)^{12} - \left( \frac{\sigma}{r_{c}} \right)^6 \right)
\end{equation}
for $r_{ij}<r_c$ and $V_{ij}=0$ for $r_{ij}>r_c$. Here $\epsilon$ sets the energy scale, and we take the range of the non-bonded interactions to be $r_c=2.5$. For the bonded interactions we use a very stiff harmonic  potential,
\begin{equation}
V_{ij}^{b}=\frac{k}{2}\left(r_{ij}-\sigma\right)^2,
\end{equation}
where $k=2000\epsilon/\sigma^2$. In this paper we report results in reduced LJ simulation units, e.g., the temperature $T=kT^*/\epsilon$ and the time $\tau=\tau^*\sqrt{\epsilon/m\sigma^2}$, where $T^*$ and $t^*$ are defined in laboratory units and $m$ is the mass of a particle. For this model the bulk glass transition temperature is estimated to be $T_g=0.44\pm 0.01$, arrived at by measuring the specific volume during constant-cooling-rate simulations from above to below $T_g$. All of our simulations were run in the NVT ensemble at with a timestep of $\delta t = 0.001\tau$ using the HOOMD-blue simulation package \cite{hoomd-blue0,hoomd-blue}.

The films were prepared by first pre-packing non-interacting random walks with the correct single-chain statistics in a simulation cell with the dimensions of the desired film \cite{Auhl2003,Sussman2016}. By treating two of the three dimensions as periodic and the other ($z$-) dimension as a reflecting wall at this step we are able to start with a configuration of chains that will not be greatly perturbed when the $z$ direction of the box is expanded to expose the surfaces of the film to vacuum. We then use a collection of chain-altering Monte Carlo (MC) moves that simultaneously reduce the density fluctuations of the pre-packed chains and respect the single-chain statistics imposed by the model potentials and by the thin film geometry. From these non-interacting chain configurations the box was expanded in the $z$ direction to allow for a free-standing film, the LJ and bonded interactions were slowly turned on, and molecular dynamics using the potentials and parameters above were begun. All configurations were prepared at an initial reduced temperature of $T=0.8$ and cooled at a constant rate of $\Gamma = 1\times 10^{-3}$ to the desired temperature; the aging time $t_{ag}$ refers to simulation time since the end of this cooling procedure. For the supported films, we first took the configuration of a thick ($\sim 10\sigma$) slab of particles from the center of a freestanding film at $T=0.425$ and $t_{ag}\approx10^6\tau$. We permanently fixed the positions of these particles and then brought an equilibrated freestanding film at $T=0.8$ in contact with this frozen layer, and then repeated the cooling procedure described above.

\subsection{Definitions of structural features}
Here we detail the definitions of the structure functions $G_{Q_l}$ and $G_{Y_{lm}}$ used in the main text. For the first we follow Steinhardt et al. \cite{Steinhardt1983} and define the function as:
\begin{equation}
G_{Q_l}(i;r_{min},r_{max}) = \left( \frac{4\pi}{2l+1}\sum_{m=-l}^l  \left| \langle  Q_{lm}(\vec{r}) \rangle  \right|^2    \right)^{1/2},
\end{equation}
where
\begin{equation}
\langle  Q_{lm}(\vec{r}) \rangle  = \frac{1}{N_j} \sum_j Y_{lm} (\hat{R}_{ij}), \  \ \ \  r_{min} < R_{ij} < r_{max}.
\end{equation}
From the above it is clear that $\langle  Q_{lm}(\vec{r}) \rangle$ is the average value of the spherical harmonics, $Y_{lm}(\vec{r})$, for every particle $j$ in a shell near particle $i$, and $G_{Q_l}(i;r_{min},r_{max})$ forms a rotationally invariant combination of the $Q_{lm}$'s. For our structure functions we choose $r_{min}$ and $r_{max}$ to be shells of width $0.5 \sigma$ where the inner radius starts at 1.0, 1.5, and 2.0 (all in units of $\sigma$). We chose $l$ parameters in the set $l \in \left\{  2,4,6,8,10,12,14\right\}$.

To define the $G_{Y_{lm}}$ structure functions we directly compute spherical harmonics using $z$ as the special axis of symmetry:
\begin{equation}
G_{Y_{lm}}(i;l;m;r_{min},r_{max}) =\frac{1}{N_j}\sum_j Y_{lm}(\hat{R}_{ij}),
\end{equation}
where for convenience we work in the basis where the $Y_{lm}$'s themselves are real. We again use the set  $l \in \left\{  2,4,6,8,10,12,14\right\}$, and consider all $-l < m < l$.

We note that, in addition to the machine learning tests discussed in the main text, we also attempted more intricate tests by expanding our set of labels to (1) $|z| < 5\sigma$, $p_h>p_c$, (2) $|z | < 5\sigma$, $p_h<p_{c,l}$,  (3) $5\sigma < |z | < 10\sigma$, $p_h>p_c$, and (4) $5\sigma < |z | < 10\sigma$, $p_h<p_{c,l}$. That is, we took pre-identified ``soft'' and ``hard'' particles from different sections of the film. Using the same generalized set of structure functions, the difference between any pair hard and soft populations is easily detectable.  On the other hand, trying to distinguish the position-based labels from each other leads to very low but positive predictive accuracy (cross-validation accuracies of $\sim 54$\% were obtained). This is consistent with the idea that in different layers of the film particles above and below fixed $p_h$ cutoffs corresponds to investigating particles with different mean softness cutoffs. Thus, this weak predictive accuracy is consistent with our finding that the structure itself is the same throughout the film.

\subsubsection{Classification with engineered and raw features}
\begin{figure}[h]
\centerline{\includegraphics[width=0.95\textwidth]{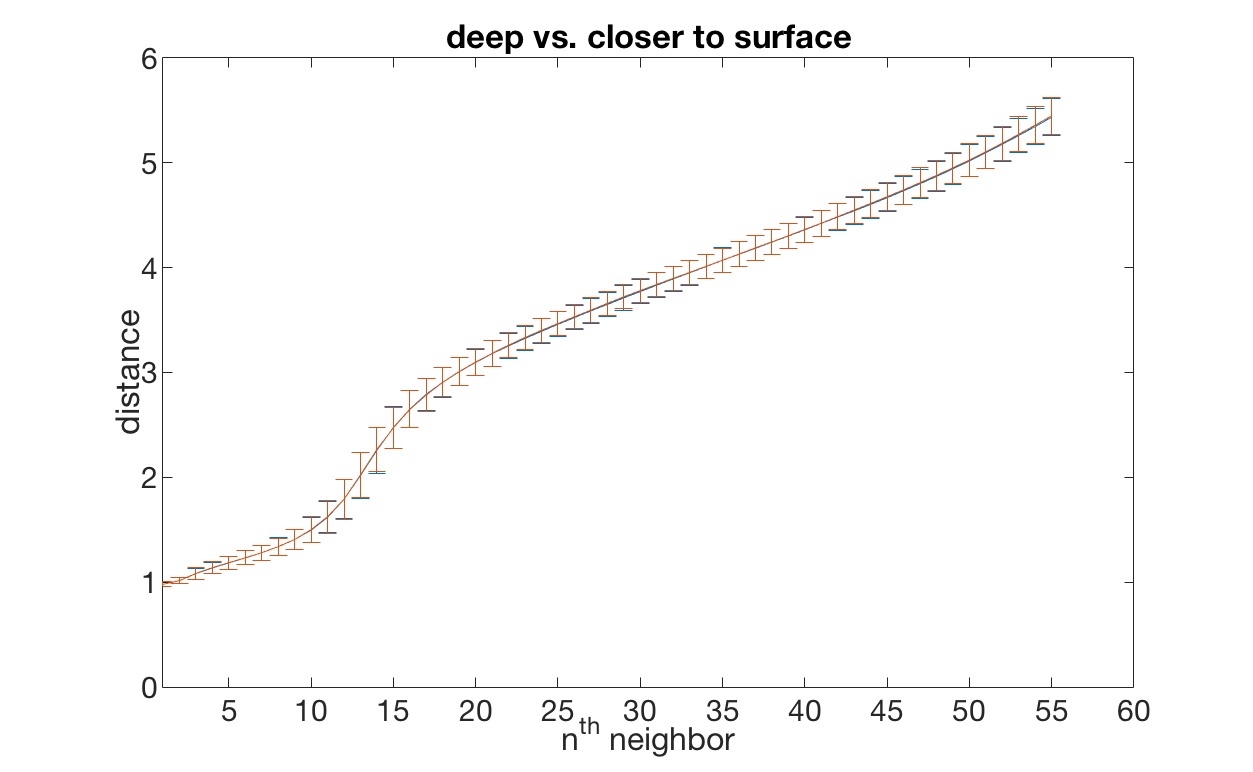}}
\caption{\label{fig:dogusSup} 
Mean distances and variances of the $n^{\textrm{th}}$ nearest neighbor of a given particle for particles near the film center (blue) and closer to the edge (red).} 
\end{figure}

We have applied several classification algorithms on the ``center'' vs. ``edge'' data sets. This includes support vector machines with linear and radial basis function kernels, as well as feedforward neural networks with up to six fully connected layers. We applied these classification algorithms using both the symmetry functions described above and in the main text, as well as with a very ``raw'' set of structural descriptors, $n(i)_{j}$, corresponding to distance of the $n^{\textrm{th}}$ nearest neighbor of particle $i$. This raw set of descriptors was shown to work when classifying ``soft'' and ``hard'' particles drawn from bulk systems. The mean distances and their standard deviations are shown in Fig. \ref{fig:dogusSup} for center and edge particles, in blue and red, respectively. We have also confirmed that adding features such as the neighbor distances along $z$ vs. in the $x-y$ plane does not give us any predictive power.
\onecolumngrid

\twocolumngrid

\subsection{Training on dynamical features}

\begin{figure}[htbp]
\centerline{\includegraphics[width=0.4\textwidth]{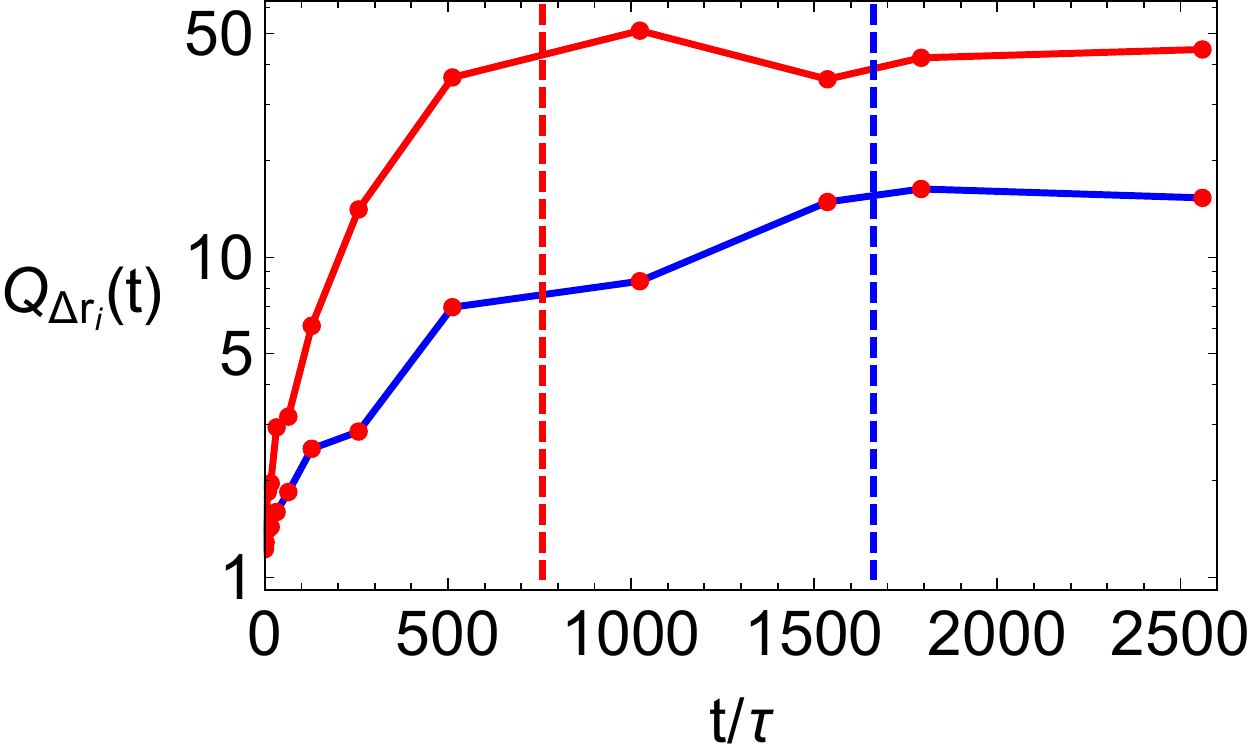}}
\caption{\label{fig:qvtm} 
$Q$ function scores obtained trying to distinguish 2500 particles near the center of the film vs particles with $12\sigma < |z-z_{cm}| < 13\sigma$ (red, upper curve) and vs particles with  $9.75\sigma < |z-z_{cm}| < 10.75\sigma$ (blue, lower curve). Vertical dashed lines are proportional to the $\tau_\alpha$ as measured by the decay of the overlap function.}
\end{figure}
Our definition of the dynamical structure function $\Delta r_{i}(t)$, is the set of instantaneously measured displacement magnitudes, $\left\{ |r_i(0) - r_i(t/\tau)| \right\}$ for the times less than $t$ in the set of approximately logarithmically spaced data, $t/\tau\in \left\{0, 2, 4, 8, 16, 32, 64, 128, 256, 512, 1024, 1536, 1792, 2560  \right\}$. Aside from the details of the time intervals considered, we note that this is a subset of the data that would be used to compute the alpha relaxation time from the decay of the overlap function \cite{Keys2007},
\begin{equation}
q(t) = 1/N_z \sum_i \Theta \left( \sigma/2 - |\mathbf{r}_i(t)-\mathbf{r}_i(0)| \right).
\end{equation}
The relaxation time is defined as a particular transformation (a thresholding) of the data, and the machine learning algorithm corresponds to a different nonlinear transformation of the data. Thus, one might expect to be able to detect the same time scales by machine learning, namely $\tau_\alpha$ for the different layers being compared. This indeed seems to be the case. The three data points in Fig. 3 of the main text rise dramatically with increasing time, but one does not expect this rapid growth to continue indefinitely. Figure \ref{fig:qvtm} shows the $Q$ score for distinguishing edge from central particles in the $T=0.425$, $t_{ag}=2\times 10^5$ thin films as a function of the time used in $\Delta r_{t}$ and the position of the edge particle population. Our data are insufficient to see the longer time scale of the central particles, but we see that for two different layers there is a crossover in the behavior of $Q$ at a time proportional to the relaxation time in each layer of the film.

\subsection{Probability of rearrangements}
In addition to the $Q$ function used in the main text, one can directly calculate the relationship between the probability of a particle rearranging ($p_h > p_{c}$)  and its softness for different layers in the film. In Ref. \cite{Schoenholz2016} we showed that the probability of rearrangements, $P_R$, could be accurately expressed as $P_R(S) = P_0(S) \exp\left(-\Delta E(S)/T\right)$; this was confirmed in bulk systems by the collapse of $P_R(S)/P_0(S)$ when plotted against $\Delta E(S)/T$ for many temperatures. The energy scale was shown to depend nearly linearly on $S$, $\Delta E = e_0 - e_1 S$, and the prefactor $P_0(S)= \exp(\Sigma (S))$ varied with $\Sigma = \Sigma_0 - \Sigma_1 S$.

\begin{figure}[htbp]
\centerline{\includegraphics[width=0.4\textwidth]{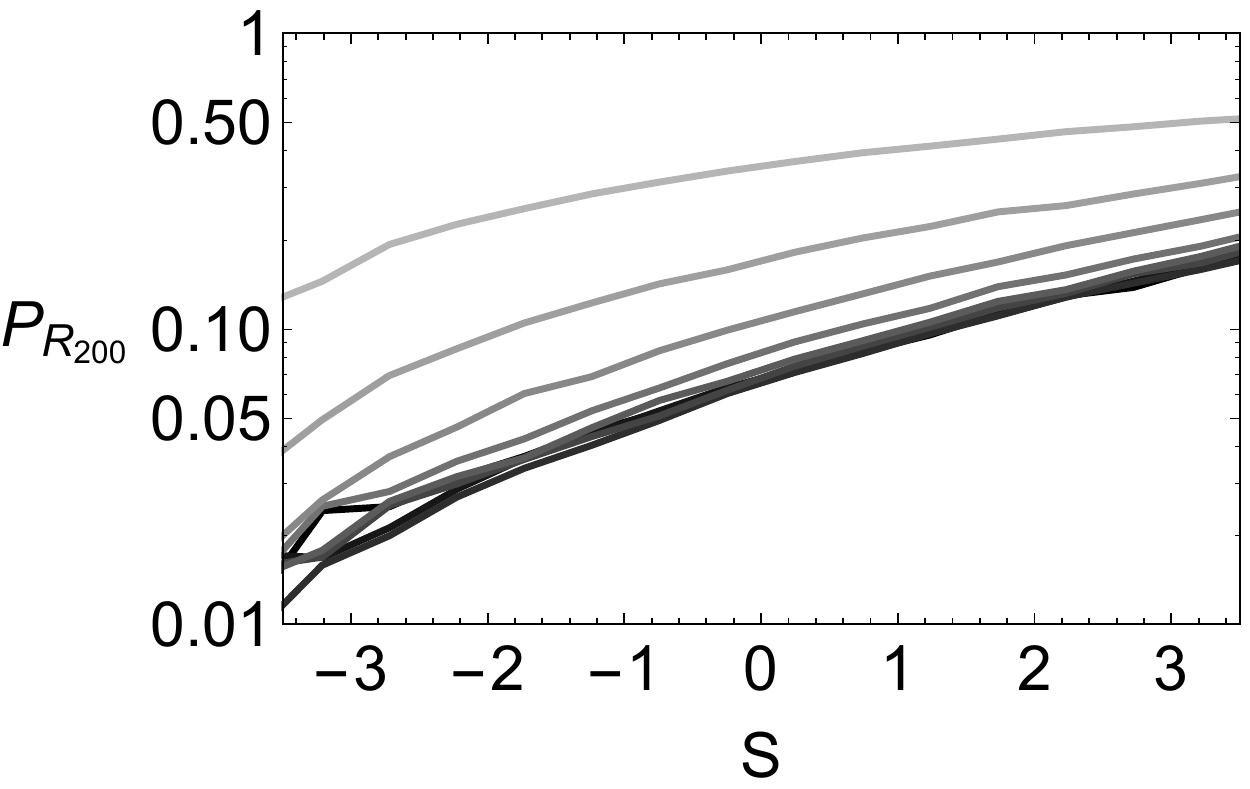}}
\caption{\label{fig:prvs} 
Probability of rearrangement measured using $dt = 200\tau$-spaced data as a function of softness for different layers of a $T=0.425$, $t_{ag} \approx 5 \times 10^5$ film. The curves are averages over particles with (black to grey) $|z-z_{cm}| < 1.5\sigma$, $1.5\sigma < |z-z_{cm}| < 3\sigma,\ldots, 12\sigma < |z-z_{cm}| < 13.5\sigma$.}
\end{figure}

In Fig. \ref{fig:prvs} we report the $P_R(S)$ for different layers in the film, again using the configuration with $T=0.425$ and $t_{ag} \approx 5 \times 10^5$. Note that on the logarithmic scale, the results for different layers are related by nearly vertical shifts. This indicates that, to first order, the behavior of the film can be captured in the same framework as the bulk systems but with an increasing value of the non-structural contribution to mobility, $\Sigma_0-e_0/T$.

\subsection{Softness relaxation}
\begin{figure}[htbp]
\centerline{\includegraphics[width=0.49\textwidth]{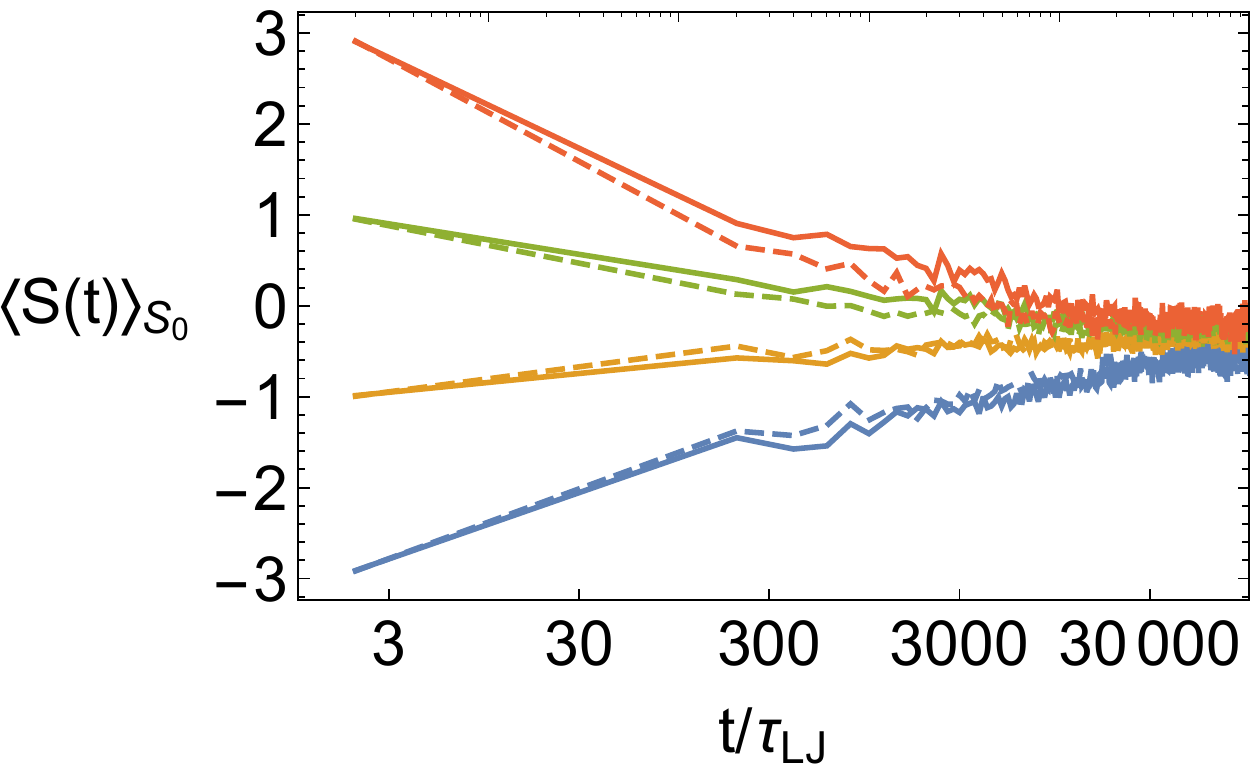}}
\caption{\label{fig:svt} 
(left) Time evolution of softness for paticles, in bins of softness, for different layers of a $T=0.425$, $t_{ag} \approx 5 \times 10^5$ film. Solid lines correspond to averages over particles with $|z-z_{cm}|<4\sigma$, and dashed lines correspond to particles in the $8\sigma<|z-z_{cm}|<12\sigma$. At $t=0$ all particles whose softness values is $\pm 1.0$ of $S_0$ are included in the average.}
\end{figure}

Earlier we noted that the ``softness propagator,'' $G(S,S_0,t)$, describes the time evolution of particles starting with softness $S_0$ and allows for predictions of the relaxation time of bulk systems to be made. One way of characterizing how softness evolves is to compute averages over this propagator
\begin{equation}
\langle S(t) \rangle_{S_0} = \int SG(S,S_0,t)\ dS.
\end{equation}
The softness of these particles can change both by rearranging themselves or by nearby particles rearranginge. In bulk KA systems we found that for each $S_0$ the average softness of particles evolves towards the mean of the softness distribution on approximately the scale of the $\alpha$-relaxation time \cite{Schoenholz2016}. Here we compute this quantity in a layer-resolved way, with the results shown in Fig. \ref{fig:svt}. To collect sufficient statistics we average over particles we define ``center'' and ``edge'' particles a bit more broadly, taking slices of the film of width $4\sigma$. While there are modest differences in how quickly softness relaxes towards the mean of the distribution, these differences are not nearly as large as the (roughly factor of two) difference in the mean relaxation times of these two populations of particles.  Thus, the enhanced dynamics leave unperturbed not only the structure, but even the time-evolution of the structure due to self- nearby rearrangements.

\end{document}